# Features

## Low Power Artificial Neural Network Architecture

Krishna Prasad Gnawali, *Student Member, IEEE,* Seyed Nima Mozaffari, *Student Member, IEEE*,
and Spyros Tragoudas , *Senior Member, IEEE*
Department of Electrical and Computer Engineering, Southern Illinois University Carbondale, USA

**Abstract** –Recent artificial neural network architectures improve performance and power dissipation by leveraging resistive devices to store and multiply synaptic weights with input data. Negative and positive synaptic weights are stored on the memristors of a reconfigurable crossbar array (MCA). Existing MCA-based neural network architectures use high power consuming voltage converters or operational amplifiers to generate the total synaptic current through each column of the crossbar array.  This paper presents a low power MCA-based feedforward neural network architecture that uses a spintronic device per pair of columns to generate the synaptic current for each neuron. It is shown experimentally that the proposed architecture dissipates significantly less power compared to existing feedforward memristive neural network architectures.

1.  **Introduction**

   Artificial neural networks are used in many applications such as pattern matching, character and speech recognition, and big data management, among others. They consist of an input layer, an output layer, and multiple hidden layers [1, 2, 3, 4]. Each layer consists of several neurons. Each neuron has multiple inputs that are typically real numbers and one output that is typically a real number. Each neuron communicates with other neurons through links called synapses that have positive or negative weight values. The neuron calculates the sum of all its weighted inputs and maps the sum into an output signal by a transfer function called activation function [1, 5-8].

   An emerging artificial neural network paradigm uses reconfigurable memristive crossbar array (MCA) to perform the needed multiplication and addition operations [2, 6, 9-13] with low power and high performance. MCA-based NN (MNN) architectures require a pair of memristors to store either a positive or a negative synaptic weight. There are different type of MNN. This paper considers multilayer feedforward MNN [2] as opposed to spiking neural networks and recurrent neural network architectures, among other types of neural networks.  The feedforward MNN in [9] uses a dual column structure where two adjacent memristors in a row store a synaptic weight. The MCA in [2] uses dual row structure where the two adjacent memristors in a column store a weight value. Both approaches store a weight value in one of the two memristors, and require the other to be in a very high resistive state so that the current through other memristor is negligible. The sign of the weight value determines which memristor is in the high resistive state. Other feedforward memristor-based NN architectures use the Wheatstone bridge [14, 15] instead of the MCA to implement a synaptic weight.  Among these approaches, the least power consuming are the dual row MCA architecture in [2] that requires a voltage converter to implement positive and negative weight values and the dual column architecture in [9] that uses an operational amplifier per column.

   Architectures as in [14-16] are gaining much attention because the required arithmetic operations can be performed by simple components that use emerging resistive devices. Power dissipation and execution time are drastically lower when compared to multiprocessor-based systems tailored to neuromorphic calculations [17, 18] or to GPGPU-based architectures [19].

   This paper presents a dual column feedforward MNN architecture that avoids the operational amplifier of [9]. Instead, it uses a spintronic device per neuron to compute the total synaptic current through each MCA column. SPICE-based simulation in 45nm technology shows that the proposed architecture dissipates considerably less power when comparing to [2] and [9]. Experimental results are presented on the benchmark data sets in [20-22].

   The paper is organized as follows. Section 2 describes the proposed architecture. Section 3 presents the experimental evaluation of the proposed architecture, and Section 4 concludes the paper.

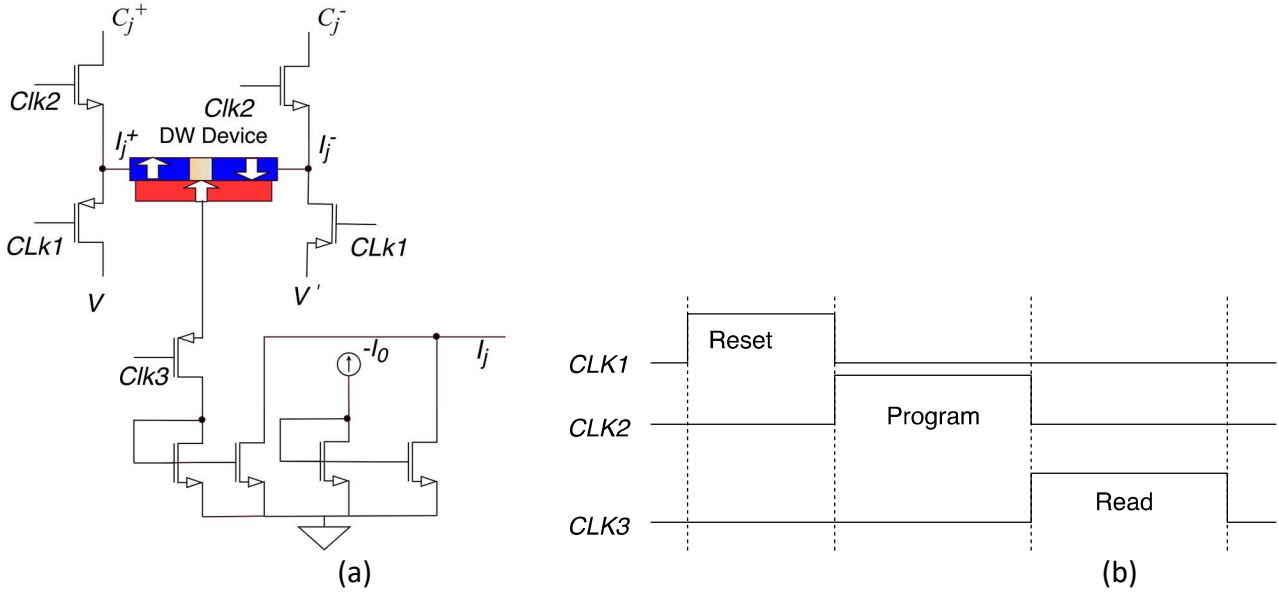

(b) Figure 2: (a) Proposed spintronic Interface module (*IM*) (b) Timing Diagram

## 2. Proposed Architecture

Figure 1 shows the structure of the proposed MCA-based layer in the feedforward MNN. It consists of *n* rows and *2m* columns. Each layer has *n* inputs $x_i$, $1 \leq i \leq n$ that are real numbers in the range [0, 1]. There are *m* neurons, and each neuron has a pair of MCA columns. At the $j^{th}$ pair, there is an interface module (denoted by IM) that generates the total synaptic current $I_j$ into the activation function module $y_j$, $1 \leq j \leq m$. The output of the $j^{th}$ activation function $y_j$, is also a real number in the range [0, 1]. Each *IM* is a domain wall spintronic device. In the current implementation, activation function $f(I_j)$ is the sigmoid and is implemented by the circuit in [23]. It is noted that the architectures in [9] and [14] only implement sigmoid activation functions. In contrast, the proposed architecture as well as [2] can accommodate any existing current-based hardware implementation of an activation function, such as the step function in [2].

Each synaptic weight $W_{i,j}$ linking the $i^{th}$ neuron and the $j^{th}$ neuron consists of two adjacent memristors $M^+_{i,j}$ and $M^-_{i,j}$, respectively. Only one of these two memristors is in the off-state. For instance, considering the $j^{th}$ column pair, if the weight is positive, $M^+_{i,j}$ is programmed to have the specific weight value, and $M^-_{i,j}$ is in the off-state. However, if the weight is negative, $M^-_{i,j}$ is programmed to have the weight value, and $M^+_{i,j}$ is in off-state. Let $I^+_j$ and $I^-_j$ denote the synaptic current for the positive and negative convolutions in the $j^{th}$ column pair.

Each column in the crossbar array of Figure 1 calculates the partial weighted sum of either positive or negative convolutions. The difference between two currents $I^+_j$ and $I^-_j$ in the $j^{th}$ column pair is calculated by the *IM*. This is the $j^{th}$ total synaptic current. Columns $C^+_j$ and $C^-_j$ in the $j^{th}$ column pair and the *IM* are part of the neuron that calculates the total synaptic current for the $j^{th}$ neuron.

Fig. 2 (a) shows the circuit diagram of the proposed *IM* where *V* denotes a control voltage. $I^+_j$ and $I^-_j$ are the inputs to the *IM* that determines the total synaptic current $I_j$. The DW device is a three-terminal device that consists of a thin nano strip between two anti-parallel fixed magnetic layers (PL). This nano strip forms the free magnetic layer (FL). The magnetization of FL determines the resistive state of the device. The transition area between the two PLs is called the DW. The DW can be moved by injecting current along the nano strip. This changes the magnetic orientation of the FL. A fixed magnet and a domain wall strip form the Magnetic Tunnel Junction (MTJ) that reads the resistive state of the device [2, 24].

The operation of the *IM* is described using three non-overlapping clocks *Clk1*, *Clk2*, and *Clk3*. The duty cycles of each clock are different because the reset, write, and read times of the DW device are different. Let $R_L$ and $R_H$ denote the low and high resistive state of device, respectively. When *Clk1* is high, the spintronic device is reset with resistive value $\frac{R_L+R_H}{2}$. In this case, the DW is at the center position. When *Clk2* is high, the spintronic device is programmed using the total synaptic current $I_j = I^+_j - I^-_j$. That way, the difference between positive and negative total synaptic current through consecutive columns of the MCA is mapped to a resistive value in the DW spintronic device. When *Clk3* is high, the activation circuit is active, and the difference of total synaptic current is mapped to a voltage value.

The current mirrors in the Figure 2 (a) ensures that the range of current generated by the interface module falls into the range required by the activation circuit for reliable operation. The externally supplied negative current $I_0$ amounts to the current through spintronic device when the DW is at the center of the nano strip. Let $I_H$ and $I_L$ denote the currents through IM when spintronic device has $R_H$ and $R_L$ resistive state, respectively. The value of $I_0 = \frac{I_H+I_L}{2}$. Figure 2 (b) shows the timing diagram of the operation of the interface module.

**Experimental Results**

A simulator for the proposed architecture has been developed. We consider $TiO_2$ bipolar metal-oxide memristors and the VTEAM model in [25]. In our simulator, the length and the $R_{ON}$, $R_{OFF}$ memristance boundaries were set to 5 nm, 5 KΩ, and 5 MΩ, respectively. Other memristor parameters were set as in [26]. The switching time was 100 ns when the applied voltage was ±1 V. Multiple bits of information can be stored in a single cell using different memristance values. Thus, $M_{i,j}^+$ and $M_{i,j}^-$ were implemented with a 5-bit memristive multi-level cell [27-29].

Since the current-voltage relation of a memristor is nonlinear, each level corresponding to a weight value was assigned using the approach presented in [29]. Any level of weight value can be realized by changing the memristance of the memristor gradually with a precise write control signal [27]. We used five different levels to implement 31 weight values. In our simulator, the dimensions of domain wall strip were $100 \times 20 \times 2$ nm³, the MgO thickness was 1.1nm, the saturation magnetization was $6.8 \times 10^5$ A/m, and domain wall width was 15 nm. The DW could be moved from one edge of free layer to the other in 2ns when applying 35 µA current.

The proposed interface module (*IM*), the voltage converter of [31], the operational amplifier-based sigmoidal neuron of [9], and the low power analog sigmoidal neuron of [23] were implemented in 45 nm predictive technology. Our experimentation showed that the average power dissipated by [23] was 8 µW.

Simulators for the MCA-based feedforward architectures in [2], [9] and non MCA-based feedforward architecture in [14] were also developed in 45 nm predictive technology for experimental comparisons. The simulator for the architecture in [2] was enhanced in order to implement the analog sigmoidal circuitry in [23]. Table 1 shows the average power dissipated by the various components in the proposed architecture as well as the MCA-based feedforward architectures in [2], [9].

Table 1 shows that the power dissipated by the proposed interface circuit was almost the same as the power of the voltage converter in the architecture of [2]. The proposed dual column architecture had reduced power when compared to the dual column architecture in [9, 16] because the number of voltage converters at any layer is equal to the number of inputs, and the number of *IM* equals to the number of neurons. The number of neurons in a layer is always less than the number of inputs.

TABLE 1
COMPONENT WISE POWER DISSIPATION

| Component | Average Power Dissipation |
|---|---|
| Proposed *IM* Module | 37 µW |
| Opamp- based Sigmoidal Neuron [9] | 104 µW |
| Voltage Converter [31] | 24 µW |

TABLE 2
COMPONENT WISE POWER DISSIPATION

| Dataset | Architecture | Average Power Dissipation (mW) | Power Reduction (%) |
|---|---|---|---|
| MNIST | Proposed | 42.10 | N.A |
| | [2] | 52.4 | 19 |
| | [9] | 97.52 | 56 |
| ASL | Proposed | 74.5 | N.A |
| | [2] | 126.2 | 41 |
| | [9] | 172.01 | 56 |
| CIFAR10 | Proposed | 37.35 | N.A |
| | [2] | 51.93 | 28 |
| | [9] | 86.32 | 56.7 |

Simulation results for the proposed architecture as well as [2] and [9] were obtained in Python for the MNIST dataset [15], American Sign Language (ASL) dataset [16], and the CIFAR10 dataset [17]. MNIST contains 28 × 28 gray scale handwritten images, ASL contains 200 × 200 RGB images, and CIFAR10 contains 32 ×32 RGB images. For the MNIST dataset, the NN had 784 input neurons and 10 output neurons. There were three hidden layers with 500, 300, 128 neurons, respectively. For the ASL dataset, the NN had 400,000 input neurons and 24 outputs. There were three hidden layers with 1,000, 500, and 128 neurons, respectively. For the CIFAR10 dataset, the NN had 1024 input neurons and 10 output neurons. There were three hidden layers with 500, 256, and 64 hidden neurons each. Images in both the ASL and CIFAR10 datasets were converted to grayscale image before feeding them into the network.

Table 2 shows the total average power dissipated by the NN architectures in [2] and [9], and the proposed architecture. The total power dissipated by the proposed architecture is the sum of the power consumed by the interface module and the sigmoidal activation function in [18]. The total power by the NN architecture in [2] amounts to the power dissipated by voltage converter and the power by the sigmoidal activation function component [23]. The total power by the NN architecture [9] amounts to the average total power dissipated by the two differential amplifiers. Notice that the power savings over [2] were approximately 19%, 41%, 28% when considering the MNIST dataset, the ASL dataset, and the CIFAR10 dataset, respectively. The power savings over the operational amplifier-based architecture in [9] was 56% for all three datasets. These results exclude the power dissipation on the MCA, which is common to all architectures.

We also provided simulation results for the NN architecture in [14]. It was simulated for a small network consisting of 10 inputs and 4 output neurons. Weights were set between 5KΩ - 5MΩ, and inputs were in the range [0, 1] V. Even for that small network, the total power dissipation was 230mW. The power dissipated by the sigmoidal neuron alone was 2.06mW. These results show that [14] is not as power efficient as the proposed architecture.

It is noted that the proposed NN architecture had the same classification accuracy as [2, 9] in all benchmarks. In particular, the accuracy was 96%, 70% and 95% for the MNIST, CIFAR10 and ASL datasets, respectively.

## 3. Conclusion

A low power spintronic circuit has been introduced in order to generate the input current for the activation circuit of an MCA-based neuromorphic architecture. The proposed interface circuit uses the domain spintronic device. It has been experimentally shown that the power dissipation of the proposed neuromorphic architecture outperforms existing architectures with emerging resistive devices.

**Reference**

[1] J. Grollier, D. Querlioz and M. D. Stiles, "Spintronic Nanodevices for Bioinspired Computing," *in Proceedings of the IEEE*, vol. 104, no. 10, pp. 2024-2039, 2016.

[2] M. Sharad, D. Fan, K. Aitken and K. Roy, "Energy-Efficient Non-Boolean Computing With Spin Neurons and Resistive Memory," *IEEE Transactions on Nanotechnology*, vol. 13, no. 1, pp. 23-34, 2014.

[3] K. Kim, J. Kim, J. Yu, J. Seo, J. Lee, and K. Choi, "Dynamic energy-accuracy trade-off using stochastic computing in deep neural networks," *ACM/EDAC/IEEE Design Automation Conference (DAC)*, pp. 1–6, 2016.

[4] Z. Li, A. Ren, J. Li, Q. Qiu, B. Yuan, J. Draper, and Y. Wang, "Structural design optimization for deep convolutional neural networks using stochastic computing," *Design, Automation Test in Europe Conference Exhibition (DATE)*, pp. 250–253, 2017.

[5] H. Chung, S. J. Lee, and J. G. Park, "Deep neural network using trainable activation functions," *International Joint Conference on Neural Networks (IJCNN)*, pp. 348–352, 2016.

[6] S. N. Mozafferi, K. P. Gnawali, S. Tragoudas, "Aging Resilient Neural Network Architecture ," *IEEE/ACM Symposium on Nanoscale Architecture (NANOARCH)*, 2018, doi: https://doi.org/10.1145/3232195.3232208.

[7] R. Kumar, S. Sharma, "Hardware Efficient Second Order Implementation of Sigmoid Function using Distributed Arithmetic," *IEEE VLSI Circuits and Systems Letter*, vol. 4, Issue-1, pp. 5-10, 2018.

[8] S. Srivastava, S. S. Rathod, "Simulation and Analysis of analog VLSI Silicon Neuron using Carbon Nanotube Field Effect Transistor and 180nm MOSFET Technology," *IEEE VLSI Circuits and Systems Letter*, vol. 3, Issue 3, pp. 37-47, 2017.

[9] R. Hasan, T. M. Taha, and C. Yakopcic, "On-chip training of memristor based deep neural networks," *International Joint Conference on Neural Networks (IJCNN)*, pp. 3527–3534, 2017.

[10] A. Basu et al., "Low-Power, Adaptive Neuromorphic Systems: Recent Progress and Future Directions," *IEEE Journal on Emerging and Selected Topics in Circuits and Systems*, vol. 8, no. 1, pp. 6-27, 2018.

[11] Chenchen Liu, Fuqiang Liu, and Hai (Helen) Li. Brain-inspired computing accelerated by memristor technology. *In Proceedings of the 4th ACM International Conference on Nanoscale Computing and Communication (NanoCom '17)*, Article 17, 6 pages, 2017.

[12] M. V. Nair, L. K. Muller, and G. Indiveri, "A differential memristive synapse circuit for on-line learning in neuromorphic computing systems," *Nano Futures*, vol. 1, no. 3, pp- 035003, 2017.

[13] G. Indiveri, B. Linares, R. Legenstein, G. Deligeorgis, T. Prodromakis, " Integration of nanoscale memristor synapses in neuromorphic computing architectures," *Nanotechnology*, volume 24, Number 38,  pp- 384010, 2013.

[14] H. Kim, M. P. Sah, C. Yang, T. Roska, and L. O. Chua, "Memristor Bridge Synapses," *Proceedings of the IEEE,* vol. 100, no. 6, pp. 2061–2070, 2012.

[15] S. P. Adhikari, H. Kim, R. K. Budhathoki, C. Yang, and L. O. Chua, "A Circuit-Based Learning Architecture for Multilayer Neural Networks With Memristor Bridge Synapses," *IEEE Transactions on Circuits and Systems I: Regular Papers*, vol. 62, no. 1, pp. 215–223, 2015.

[16] M. Prezois, F. Merrikh-Bayat, B. Hoskins, G. Adam, K. Likharev, and D. Strukov, "Training and Operation of an Integrated neuromorphic network based on Metal-Oxide Memristors," *Nature Letter*, vol. 521, no. 1, pp. 61–64, 2016.

[17] M. Davies e.t.al, "Loihi: A Neuromorphic Manycore Processor with On-Chip Learning," IEEE *Micro*, vol. 38, no. 1, pp. 82–99, 2018.

[18] F. Akopyan e.t.al "Truenorth: Design and tool flow of a 65 mw 1 million neuron programmable neurosynaptic chip," *IEEE Transactions on Computer-Aided Design of Integrated Circuits and Systems*, vol. 34, no. 10, pp. 1537–1557, 2015.

[19] E. Lindholm and S. Oberman, "The NVIDIA GeForce 8800 GPU," *IEEE Hot Chips 19 Symposium (HCS), Stanford, CA*, pp. 1-17, 2007.

[20] Y. LeCun, C. Cortes, and C. J. C. Burges, "The MNIST Database of Handwritten Digits," [Online]. Available: *http://yann.lecun.com/exdb/mnist/*

[21] "American Sign Language Translation," [Online]. Available: *https://github.com/mjk188/ASLTranslator/tree/master/TrainData*

[22] A. Krizhevsky, "The CIFAR10 dataset," [Online]. Available*: https://www.cs.toronto.edu/~kriz/cifar.html*

[23] G. Khodabandehloo, M. Mirhassani, and M. Ahmadi, "Analog Implementation of a Novel Resistive-Type Sigmoidal Neuron," *IEEE Transactions on Very Large Scale Integration (VLSI) Systems*, vol. 20, no. 4, pp. 750–754, 2012.

[24] K. P. Gnawali, S. N. Mozaffari and S.Tragoudas, "Low Power Spintronic Ternary Content Addressable Memory", *IEEE Transaction on Nanotechnology*, 2018, doi: 10.1109/TNANO.2018.2869734.

[25] S. Kvatinsky, M. Ramadan, E. G. Friedman, and A. Kolodny, "VTEAM: A General Model for Voltage-Controlled Memristors," *IEEE Transactions on Circuits and Systems II: Express Briefs*, vol. 62, no. 8, pp. 786–790, 2015.

[26] S. Kvatinsky, K. Talisveyberg, D. Fliter, A. Kolodny, U. C. Weiser, and E. G. Friedman, "Models of memristors for SPICE simulations*," IEEE 27th Convention of Electrical and Electronics Engineers in Israel*, pp. 1–5, 2012.

[27] F. Alibart, L. Gao, B. D. Hoskins, and D. B. Strukov, "High precision tuning of state for memristive devices by adaptable variation-tolerant algorithm," *Nanotechnology*, vol. 23, no. 7, pp. 1–7, 2012.

[28] S. Yu, Y. Wu, and H.-S. P. Wong, "High precision tuning of state for memristive devices by adaptable variation-tolerant algorithm," *Applied Physics Letter*, vol. 98, 2011.

[29] L. Zhang, D. Strukov, H. Saadeldeen, D. Fan, M. Zhang and D. Franklin, "SpongeDirectory: Flexible sparse directories utilizing multi-level memristors," *International Conference on Parallel Architecture and Compilation Techniques (PACT)*, pp. 61-73, 2014.

[30]  M. Wu, W.-Y. Jang, C. Lin, and T. Tseng, "A study on low-power, nanosecond operation and multilevel bipolar resistance switching in Ti/ZrO2/Pt nonvolatile memory with 1T1R architecture," *Semiconductor Science and Technology*, vol. 27, no. 6, p. 065010, 2012.

[31] K. I. Hwu, W. C. Tu, and Y. H. Chen, "A novel negative-output KY boost converter," *International Conference on Power Electronics and Drive Systems (PEDS)*, pp. 1155–1157, 2009.

5## About the Authors

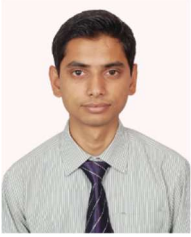

**Krishna Prasad Gnawali** (krishna.gnawali@siu.edu) received his bachelors degree in Electronics and Communication Engineering from Tribhuvan University, Nepal in 2014. Since 2016, he is a Ph.D. student in the Department of Electrical and Computer Engineering at the Southern Illinois University Carbondale. He is a member of the Electronic Design & Test Automation Lab, where he is searching for design and test of emerging technologies.

His research interests include VLSI Design and Test Automation, RTL-level test and verification, Neuromorphic computing, emerging technologies (memristors, spintronics), and Logic in memory processing.

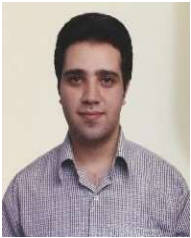

**Seyed Nima Mozaffari** (nima.mozaffari@siu.edu) received his M.S. degree from School of Electrical and Computer Engineering at the University of Tehran in 2010, and PhD degree in Electrical and Computer Engineering in 2018 from Southern Illinois University Carbondale. His research interests include VLSI Design and Test Automation, RTL-level test and verification, Neural network and neuromorphic computing, emerging technologies (memristors, spintronics), aging-aware design and analysis Synthesis of Threshold Logic Gates, and Synthesis on non-boolean near threshold functions. He is a reviewer in IEEE journals and conferences

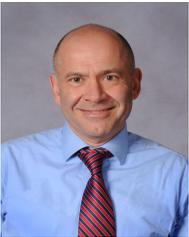

**Spyros Tragoudas** (spyors@siu.edu) received his Diploma in Computer Engineering from the University of Patras, Greece in 1986, and M.S. and PhD degrees in Computer Science from the University of Texas at Dallas in 1988 and 1991. He is Professor and Department Chair at the Electrical and Computer Engineering (ECE) Department, Southern Illinois University at Carbondale (SIUC), and the Director of the National Science Foundation (NSF) Industry University Cooperative Research Center (IUCRC) on Embedded Systems at the SIUC site. He has held prior appointments with the faculty of the ECE Department at the University of Arizona, and with the faculty of the Computer Science Dept. at SIUC.

His current research interests are in the areas of VLSI design/test automation and embedded systems. Dr. Tragoudas has published over two hundred and fifty papers in peer-reviewed journals and conference proceedings as well as book chapters in these areas. He has received three outstanding paper awards, and has directed many PhD dissertations and MS theses. His research has been funded by federal agencies and industry. He has served and current serving on the editorial board of several journals, the technical program committees of many conferences.